\documentclass[a4paper,10pt]{article}

\usepackage{amsmath}

\usepackage[english]{babel} 
\usepackage{graphicx} 
\usepackage{multirow} 
\usepackage[center]{caption}
\usepackage{geometry}
\usepackage{indentfirst}
\usepackage{txfonts}

\graphicspath{{Figures_artISMA2014/}}

\usepackage{cases}
\usepackage{subfigure}
\usepackage[table]{xcolor}
\usepackage[utf8x]{inputenc}
\usepackage[T1]{fontenc}
\usepackage{color}
\usepackage{abstract}

\usepackage{hyperref}
\hypersetup{
    colorlinks,%
    citecolor=red,%
    filecolor=black,%
    linkcolor=blue,%
    urlcolor=black
}

\title{\bfseries Effect of the shape of mouth pressure variation on dynamic oscillation threshold of a clarinet model}

\author{B. Bergeot$^{a,b,*}$ , A. Almeida$^{b,c}$ and C. Vergez$^{a}$\vspace{12pt}\\
{\small$^{a}$LMA, CNRS UPR7051, Aix-Marseille Univ., Centrale Marseille, F-13402 Marseille Cedex 20, France}\\
 {\footnotesize$^{b}$LUNAM Universit\'{e}, Universit\'{e} du Maine, UMR CNRS 6613, Laboratoire d’Acoustique, Avenue Olivier Messiaen, 72085 Le Mans Cedex 9, France}\\
{\footnotesize$^{c}$ School of Physics, The University of New South Wales, Sydney UNSW 2052, Australia}
}

\date{\itshape Proceeding of the International Symposium on Musical Acoustics (ISMA)\\
July 7-12, 2014, Le Mans France}

\begin{document}

\twocolumn[
\begin{@twocolumnfalse}

\maketitle

\begin{abstract}
\noindent Simple models of clarinet instruments based on iterated maps have been used in the past to successfully estimate the threshold of oscillation of this instrument as a function of a constant blowing pressure. However, when the blowing pressure gradually increases through time, the oscillations appear at a much higher value, called dynamic oscillation threshold, than what is predicted in the static case.

\noindent This is known as bifurcation delay, a phenomenon studied in~\cite{BergeotNLD2012,BergeotNLD2012b} for a clarinet model. In particular the dynamic oscillation threshold is predicted analytically when the blowing pressure is linearly increased. However, the mouth pressure cannot grow indefinitely. During a note attack, after an increasing phase, the musician stabilizes the mouth pressure. In the present work, the analytical prediction of the dynamic oscillation threshold is extended to a situations in which the mouth pressure approaches a steady state pressure according to an exponential time profile. The predictions still show a good agreement with simulation of the simple clarinet-model. This situation is   compared in terms of dynamic oscillation bifurcation. 
\end{abstract}
\vspace{24pt}
\end{@twocolumnfalse}]


\section{Introduction}
\label{sec:introduction}

One of the main sucesses of clarinet models, even if extremely simplified, is that they can predict ranges of parameters such as blowing pressure and lip force where the instrument produces a sound. The oscillation threshold, i.~e. the minimum blowing pressure at which there can be a sustained oscillation has been extensively studied \cite{KergoActa2000,dalmont:3294}. The oscillation threshold can be measured by applying a constant blowing pressure, allowing enough time to let the system reach a permanent regime (either non-oscillating or strictly periodic), and repeating the procedure for other constant blowing pressures. This is a static or stationary view of the threshold.

A model based on an iterated map can be used to predict the asymptotic (or \textit{static}) behavior (in particular the static oscillation threshold) of a simplified clarinet model as a function of a constant mouth pressure. This procedure avoids the phenomenon of \textit{bifurcation delay}~\cite{Fruchard2007}, a shift of the pressure at which the oscillation starts when the pressure is gradually increased over time.  In such a situation, the value of blowing pressure at which sound is observed is called \textit{dynamic oscillation threshold}, in opposition to the \textit{static oscillation threshold}. The phenomenon has been observed by Bergeot \emph{et al.} in numerical simulations~\cite{BergeotNLD2012} and experiments~\cite{Jasa2013BBergeot}. Using a simplified clarinet model the dynamic oscillation threshold has also been predicted \cite{BergeotNLD2012,BergeotNLD2012b} when the mouth pressure is linearly increased.

In most realistic situations (for example during a note attack) the pressure stabilises at a target value as the oscillations grow to an audible level. The present paper shows how analytical results of \cite{BergeotNLD2012,BergeotNLD2012b} can be extended to predict the dynamic oscillation threshold of the system for an archetypal mouth pressure shape that smoothly approaches the target value exponentially. The mathematical procedure and a comparison of theoretical results with numerical simulations are presented in section~\ref{sec:ISMAsec3}.  Then, in section~\ref{sec:4}, a preliminary investigation on the influence of the mouth pressure shape on the onset of oscillations is performed. The clarinet model and major results  from~\cite{BergeotNLD2012,BergeotNLD2012b} are first briefly recalled in section~\ref{sec:STA}.

\section{State of the art}\label{sec:STA}

\subsection{Clarinet Model}\label{sec:ClarMod}

This model divides the instrument into two elements: the exciter and the resonator. The exciter is represented by a nonlinear function $F$ also called nonlinear characteristic of the exciter, relating the pressure applied to the reed $p(t)$ to the flow $u(t)$ through its opening. The resonator (the bore of the instrument) is described by its reflection function $r(t)$. $p$ and $u$ are two non-dimensional state variables that are sufficient to describe the state of the instrument.

The solutions $p(t)$ and $u(t)$ depend on the control parameters: $\gamma$ proportional to the mouth pressure $P_m$ according to
\begin{equation}
  \label{eq:1}
  \gamma = \frac{P_m}{P_M} = \frac{P_m}{kH}
\end{equation}
where $P_M=kH$ represents the pressure needed to close the reed entrance (also used to normalize the pressure $p(t)$), where $1/k$ is the acoustic compliance of the reed and $H$ its distance to the lay at rest. The other parameter is $\zeta$ which is related to the opening of the embouchure at rest according to the formula
\begin{equation}
  \label{eq:2}
  \zeta ={Z_c \,U_A}/{P_M}=Z_c w H \sqrt{\frac{2}{\rho P_M}} .
\end{equation}
Here, $Z_c$ is the characteristic impedance at the input of the bore, $w$ the effective width of the reed, and $U_A$ the maximum flow admitted by the reed valve. Biting harder the embouchure reduces the value of $\zeta$. For most of the analysis below, this parameter is maintained at $0.5$, but the analysis can easily be reproduced for other values of $\zeta$. The nonlinear characteristic is provided by the Bernoulli equation describing the flow in the reed channel plus some additional hypothesis on the turbulent mixing within the mouthpiece~\cite{HirschAal1990,MOMIchap7}. 

The model is extremely simplified by considering a straight resonator in which the eventual losses are independent of frequency. In the current work, losses are ignored in all calculations. The reed is considered as an ideal spring \cite{Maga1986,MechOfMusInst,OllivActAc2005,dalmont:3294,Cha08Belin}. With these assumptions, the reflection function becomes a simple delay with sign inversion. Using the variables $p^+$ and $p^-$ (outgoing and incoming pressure waves respectively) instead of the variables $p$ and $u$, the system can be simply described by an iterated map \cite{Maga1986}:

\begin{equation}
p^+_n=G\left(p^+_{n-1},\gamma \right).
\label{Def_stat}
\end{equation}

Function $G$ can be determined from the nonlinear characteristic $F$, which is done by Taillard \cite{NonLin_Tail_2010} for $\zeta<1$. This function depends on the control parameters $\gamma$ and $\zeta$. The time step $n$ corresponds to the round trip time $\tau=2l/c$ of the wave with velocity $c$ along the resonator of length $l$.

Using the universal properties of iterated maps \cite{Feigen1979},  useful information about the instrument behavior can be drawn from the study of the iteration function. Most of these studies are done in the context of \textit{static} bifurcation theory, which assumes that the control parameter $\gamma$ is constant. For instance, it is possible to determine the steady state of the system as a function of the parameter $\gamma$, and to plot a bifurcation diagram shown in red in Fig.~\ref{fig12}, in terms of variable $p^+$. When no losses are considered, we have $\gamma_{st}=1/3$. For all values of the control parameter $\gamma$ below $\gamma_{st}$ the series $p^+_n$ converges to a single value $p^{+*}$ corresponding to the fixed point of the function $G$, i.e.~the solution of  $p^{+*}=G\left(p^{+*}\right)$. When the control parameter $\gamma$ exceeds $\gamma_{st}$ the fixed point of $G$ becomes unstable and the steady state becomes a 2-valued oscillating regime. 

\subsection{Slowly linear time-varying mouth pressure}\label{sec:2:2}

\subsubsection{Dynamic bifurcation}

A control parameter $\gamma$ increasing linearly with time is taken into account by replacing Eq.~(\ref{Def_stat}) by Eqs.~(\ref{dynsys_pp_a}) and (\ref{dynsys_pp_b}):

\begin{subnumcases}{\label{dynsys_pp}}
p^+_n=G\left(p^+_{n-1},\gamma_n\right)\label{dynsys_pp_a}\\
\gamma_{n}=\gamma _{n-1} +\epsilon.\label{dynsys_pp_b}
\end{subnumcases}
 
\begin{figure}[t]
\centering
\includegraphics[width=1\columnwidth]{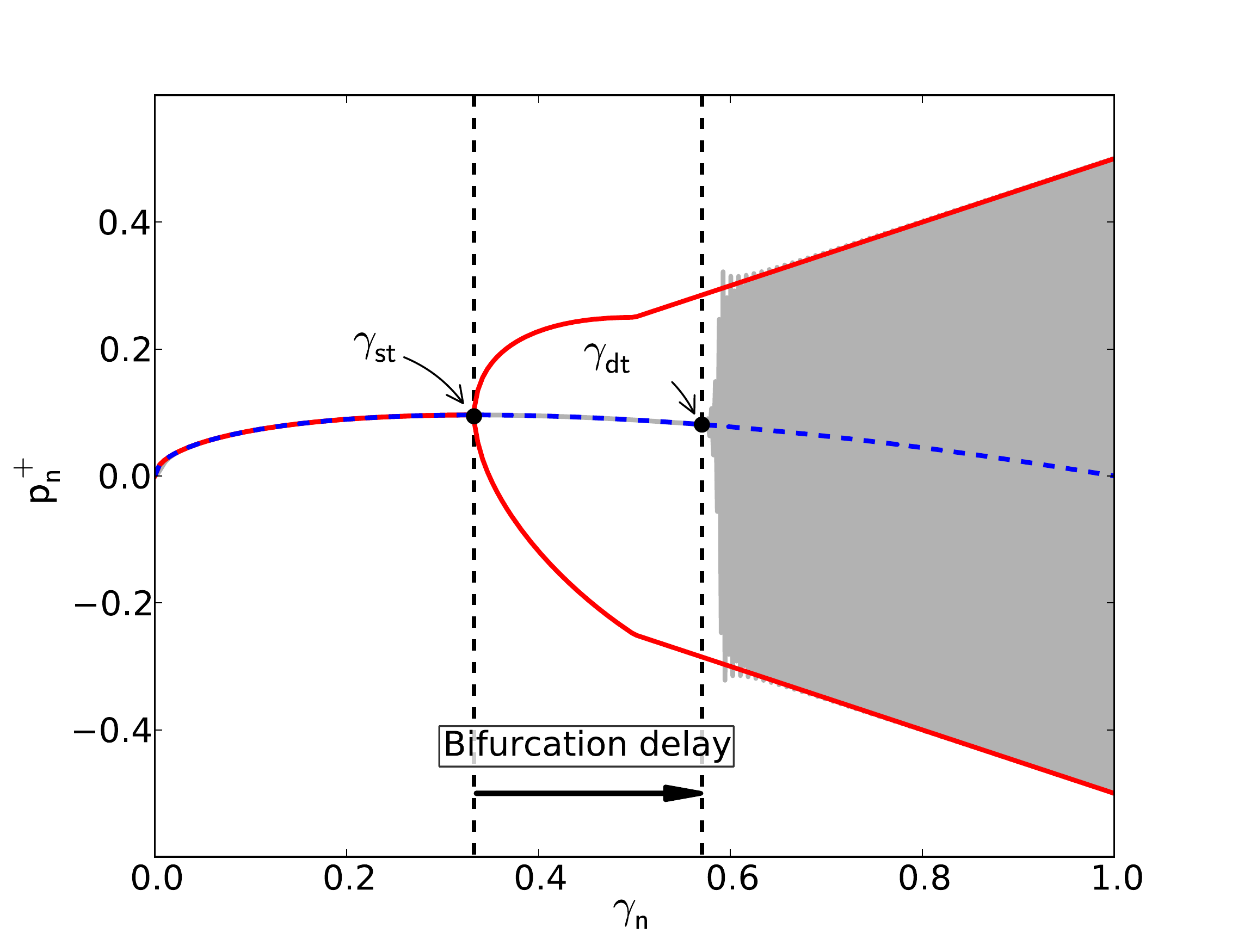}
\caption{Comparison between \textit{static} and \textit{dynamic} bifurcation diagram as functions of $\gamma_n$. $\epsilon=2\cdot10^{-3}$, $\zeta=0.5$. The phenomenon of bifurcation delay is highlighted.}
\label{fig12}
\end{figure}

$\gamma$ is assumed to increase slowly, hence $\epsilon$ is considered arbitrarily small ($\epsilon\ll 1$). When the series $p^+_n$ is plotted with respect to parameter $\gamma_n$ the resulting curve can be interpreted as a \textit{dynamic} bifurcation diagram and it can be compared to the \textit{static} bifurcation diagram (Fig.~\ref{fig12}).

Because of the time variation of $\gamma$, the system in Eqs.~(\ref{dynsys_pp}) is subject to the phenomenon of bifurcation delay \cite{Baesens1991,Fruchard2007}: the bifurcation point (in this case the oscillation threshold) is shifted from the \textit{static oscillation threshold} $\gamma_{st}$~\cite{dalmont:3294} to the \textit{dynamic oscillation threshold} $\gamma_{dt}$~\cite{BergeotNLD2012}. The difference $\gamma_{dt}-\gamma_{st}$ is called the \textit{bifurcation delay}. In a previous work \cite{BergeotNLD2012}, the bifurcation delay was found to depend very strongly on noise, in particular due to round-off errors made by the computer in numerical simulations, even if high precisions were used.

According to dynamic bifurcation theory, two operative regimes must be distinguished~ \cite{Baesens1991}:  

\begin{itemize}
\item The \textbf{Deterministic Regime (DReg.):} In this case, the noise does not affect the bifurcation delay which does not depend on the slope $\epsilon$ of the blowing pressure.
\item The \textbf{Sweep-Dominant Regime (SDReg.):} In this case, the bifurcation delay is affected by the noise, becoming larger as the blowing pressure $\gamma$ is increased quicker.
\end{itemize}

Articles~\cite{BergeotNLD2012,BergeotNLD2012b} provide an analytical study of the dynamic bifurcation of the clarinet model (i.e.~Syst.~(\ref{dynsys_pp})) based on a generic method given by Baesens~\cite{Baesens1991}. The main results of these studies are theoretical estimations of the dynamic oscillation threshold of the clarinet: one for the DReg.~\cite{BergeotNLD2012} and one for the SDReg.~\cite{BergeotNLD2012b}. These expression are recalled below.

\subsubsection{Dynamic oscillation threshold for the deterministic regime}
\label{sssec:DyThDet}

A possible theoretical estimation of the dynamic oscillation threshold consists in identifying the value of $\gamma$ for which the orbit of the series $p^+_n$ escapes from a neighborhood of arbitrary distance of an \textit{invariant curve} $\phi(\gamma,\epsilon)$. More precisely, the dynamic oscillation threshold is reached when the distance between the orbit and the invariant curve becomes equal to $\epsilon$.

The invariant curve (i.e. invariant under the mapping (\ref{dynsys_pp}), described for example in \cite{FruchaScaf2003}) can be seen as the equivalent of a fixed point in static regimes. It satisfies the following equation: 
\begin{equation}
\phi_\epsilon(\gamma)=G\left(\phi_\epsilon(\gamma-\epsilon),\gamma\right).
\label{eqdiff_1}
\end{equation}

Although Eq.~\eqref{eqdiff_1} usually leads to mathematical expressions that cannot be calculated analytically, the invariant curve can be determined approximately with a perturbation method given by Baesens~\cite{Baesens1991}, leading to the following general form:

\begin{equation}
\phi_\epsilon(\gamma)=\sum_{i=0}^n \epsilon^i \phi_i(\gamma) + o(\epsilon^{n+1}),
\end{equation}
where the zeroth order term of the series is the fixed point curve of the function $G$, $\phi_0(\gamma)=p^{+*}(\gamma)$.

The procedure to obtain the theoretical estimation $\gamma_{dt}^{th}$ of the dynamic oscillation threshold is as follows: a theoretical expression of the invariant curve is found for a particular (small) value of the increase rate $\epsilon$ (i.e. $\epsilon \ll1$). Equations~(\ref{dynsys_pp}) are then expanded into a first-order Taylor series around the invariant curve and the resulting linear system is solved analytically. Finally, $\gamma_{dt}^{th}$ is derived from the analytic expression of the orbit.

The analytic estimation of the dynamic oscillation threshold $\gamma_{dt}^{th}$ is defined in~\cite{BergeotNLD2012}:
\begin{equation}
\int_{\gamma_0+\epsilon}^{\gamma_{dt}^{th}+\epsilon}\ln\left| \partial_xG\left(\phi_\epsilon(\gamma'-\epsilon),\gamma'\right)\right|d\gamma'=0,
\label{dynoscthre_2}
\end{equation}
where $\gamma_0$ is the initial value of $\gamma$ (i.e. the starting value of the linear ramp).

\subsubsection{Dynamic oscillation threshold for the sweep-dominant regime}
\label{sssec:DyThNo}

The effect of the noise can be taken into account by introducing an uniformly distributed random variable in the system described by Eqs.~\eqref{dynsys_pp}. This random variable is an additive white noise with an expected value of zero and variance $\sigma^2$. In the case of a numerical simulation performed with finite precision,\footnote{The precision in here referred as the number of decimal digits used by the computer.} $\sigma=10^{-precision}$.

The method to obtain the theoretical estimation of the dynamic oscillation threshold for the SDReg. (noted $\hat{\gamma}_{dt}^{th}$) is detailed in~\cite{BergeotNLD2012b}. 

Because of noise, the bifurcation delay is reduced. Therefore, the main approximation of the method is to assume that the dynamic oscillation threshold is close to the static oscillation threshold $\gamma_{st}$. Using this approximation, the expression of $\hat{\gamma}_{dt}^{th}$ is:

\begin{equation}
\hat{\gamma}_{dt}^{th} = \gamma_{st}+\sqrt{-\frac{2\epsilon}{K}\ln\left[\left(\frac{\pi}{K}\right)^{1/4}
\frac{\sigma}{\epsilon^{5/4}}\right]},
\label{eq:seuidynno}
\end{equation}
which is the theoretical estimation of the dynamic oscillation threshold for the SDReg., $K$ is a constant that depends on the slope of $\partial_x G(p^+(\gamma),\gamma)$, the derivative of the iteration function at the fixed point. 


\section{Exponential variation of the mouth pressure}
\label{sec:ISMAsec3}

During a note attack, the mouth pressure cannot grow indefinitely, being stabilized to a targeted value before the oscillations grow to an audible level. This section is devoted to show how results presented in sections~\ref{sssec:DyThDet} and \ref{sssec:DyThNo} can be extended to predict the dynamic oscillation threshold of the system for a profile in which the parameter approaches asymptotically a target value through an exponential function.

\subsection{Prediction of the dynamic oscillation threshold: mathematical procedure}

When the mouth pressure follows an exponential function, the system is described by the following system of difference equations:

\begin{subnumcases}{\label{dynsys_pp_exp}}
p^+_n=G\left(p^+_{n-1},\gamma_n\right)\label{dynsys_pp_exp_a}\\
\gamma_{n}=a \, \gamma _{n-1} +\gamma_M(1-a),\label{dynsys_pp_exp_b}
\end{subnumcases}
where $\gamma_M$ is the targeted mouth pressure (it is always equal to 1 in this work). Eq.~\eqref{dynsys_pp_exp_b} describes the exponential variation of the mouth pressure, with $\gamma_0=0$ and noting $\epsilon=-\ln [a]$, its formal solution is given by:

\begin{equation}
\gamma_n=\gamma_M \left(1-e^{-n\epsilon}\right),
\label{dynsys_pp_expfun}
\end{equation}

\begin{figure}[h!]
\centering
\includegraphics[width=1\columnwidth]{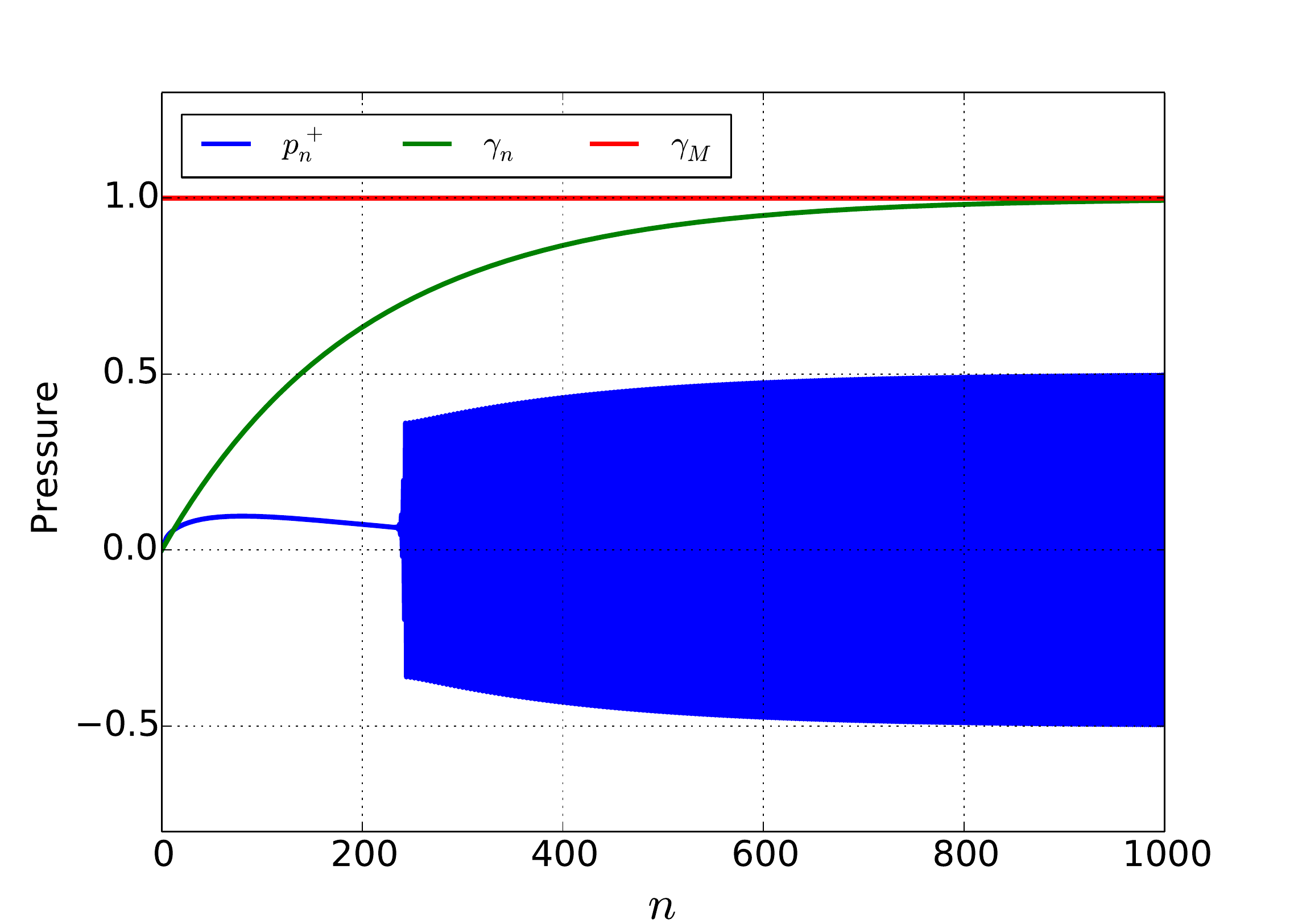}
\caption{Numerical simulation performed on the Syst.~\eqref{dynsys_pp_exp}.The parameters used are: $\gamma_M=1$, $a=0.995$ ($\epsilon=0.005$), $\zeta=0.5$, $\gamma_0=0$ and $p^+_0=G\left(0;\gamma_0\right)$.}
\label{fig:1bis}
\end{figure}

Fig.~\ref{fig:1bis} shows an example of numerical simulation performed on the system described by Eqs.~\eqref{dynsys_pp_exp}.

Using a change of parameters, Eqs.~\eqref{dynsys_pp_exp} can take the form of~\eqref{dynsys_pp}. The iterative function $G$ is replaced by $H$:
\begin{equation}
H\left(x,\eta\right)=G\left(x,\gamma(\eta)\right),
\end{equation}

 with a linearly increasing parameter $\eta$: 
\begin{equation}
\eta\left(\gamma\right)=\ln\left[\frac{\gamma_M}{\gamma_M-\gamma}\right] \; \Longrightarrow \; \gamma\left(\eta\right)=\gamma_M\left(1-e^{-\eta}\right).
\label{eq:eta_fgam}
\end{equation}

The new system is written:
\begin{subnumcases}{\label{dynsys_pp_exp2}}
p^+_n=H\left(p^+_{n-1},\eta_n\right)\label{dynsys_pp_exp2_a}\\
\eta_{n}=\eta _{n-1} +\epsilon.\label{dynsys_pp_exp2_b}
\end{subnumcases}

Using Eqs.~\eqref{dynoscthre_2} and \eqref{eq:seuidynno}, it possible to predict the dynamic oscillation thresholds, noted $\eta_{dt}^{th}$ (for DReg.) and $\hat{\eta}_{dt}^{th}$ (for SDReg.),  of the system of Eqs.~\eqref{dynsys_pp_exp2}. These thresholds are then expressed in terms of mouth pressure using Eq.~\eqref{eq:eta_fgam}: $\Gamma_{dt}^{th}=\gamma\left(\eta_{dt}^{th}\right)$ and $\hat{\Gamma}_{dt}^{th}=\gamma\left(\hat{\eta}_{dt}^{th}\right)$.

A summary table of different notations of the oscillation thresholds is provided in table \ref{ta:TaOfNotThresh}.

\begin{table}[h!]
\centering
\caption{Notation for thresholds of oscillation.}
{\small
\begin{tabular}{|p{1cm}|p{5.4cm}|}
\hline
\multicolumn{2}{ |p{6.8cm}| }{\textbf{Static oscillation thresholds}}\\ \hline
$\gamma_{st}$ & static oscillation threshold \\\hline
$\eta_{st}$ & $\eta\left(\gamma_{st}\right)$ calculated through Eq.~\eqref{eq:eta_fgam} \\\hline
\multicolumn{2}{ |p{6.8cm}|}{\textbf{Dynamic oscillation thresholds of Syst.~\eqref{dynsys_pp}}}\\ 
\multicolumn{2}{ |p{6.8cm}|}{(linear variation of the mouth pressure)}\\ \hline
$\gamma_{dt}^{th}$ & theoretical estimation of the dynamic oscillation threshold for DReg.\\\hline
$\hat{\gamma}_{dt}^{th}$ & theoretical estimation of the dynamic oscillation threshold for SDReg. \\\hline
$\gamma_{dt}^{num}$  & dynamic oscillation threshold calculated on numerical simulations \\\hline
\multicolumn{2}{ |p{6.8cm}| }{\textbf{Dynamic oscillation thresholds of Syst.~\eqref{dynsys_pp_exp2}}}\\
\multicolumn{2}{ |p{6.8cm}|}{(exponential variation of the mouth pressure)}\\ \hline
$\eta_{dt}^{th}$ & theoretical estimation of the dynamic oscillation threshold for DReg.\\\hline
$\hat{\eta}_{dt}^{th}$ & theoretical estimation of the dynamic oscillation threshold for SDReg. \\\hline
$\eta_{dt}^{num}$  & dynamic oscillation threshold calculated on numerical simulations \\\hline
$\Gamma_{dt}^{th}$, $\hat{\Gamma}_{dt}^{th}$, $\Gamma_{dt}^{num}$  & $\gamma\left(\eta_{dt}^{th}\right)$, $\gamma\left(\hat{\eta}_{dt}^{th}\right)$, $\gamma\left(\eta_{dt}^{num}\right)$ calculated through Eq.~\eqref{eq:eta_fgam} \\\hline
\end{tabular}
}
\label{ta:TaOfNotThresh}
\end{table}

\subsection{Benchmark of theoretical estimators for the dynamic threshold}

\begin{figure}[h!]
\centering
\includegraphics[width=1\columnwidth]{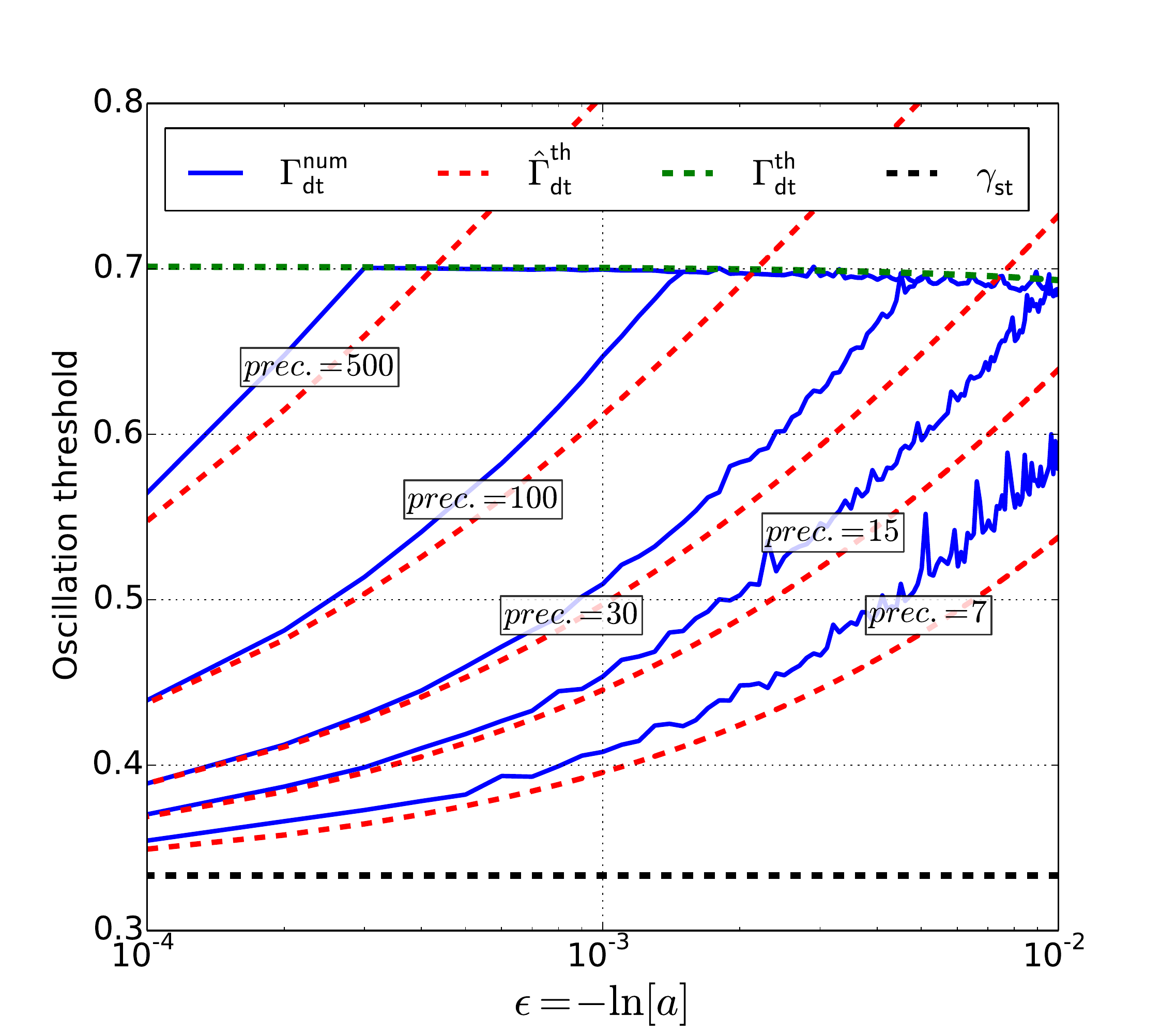}
\caption{Graphical representation of $\Gamma_{dt}^{num}$. Results are compared to analytical \textit{static} and \textit{dynamic} thresholds: $\gamma_{st}$, $\Gamma_{dt}^{th}$ and $\hat{\Gamma}_{dt}^{th}$. Different precisions are used: prec.~= 7, 15, 30, 100 and 400. $\gamma_0 = 0$ and $\gamma_M=1$.}
\label{fig:2a}
\end{figure}

In this section, the above theoretical predictions of the dynamic threshold obtained for an exponential variation of the mouth pressure are compared to numerical simulations. A numerical dynamic threshold $\eta_{dt}^{num}$ is estimated as the value for which the distance between the simulated orbit of Syst.~\eqref{dynsys_pp_exp2} and its invariant curve is first larger than $\epsilon$. The comparison is carried out as a function of parameter~$\epsilon$. Results are shown in terms of mouth pressure $\gamma$ in Fig.~\ref{fig:2a} with $\Gamma_{dt}^{num}=\gamma\left(\eta_{dt}^{num}\right)$ and for several values of the precision. The choice of the precision is possible using \emph{mpmath}, an arbitrary precision library for \textsc{Python}. 

Similarly to the case of a linear variation of the mouth pressure $\gamma$~\cite{BergeotNLD2012b}, the evolution of $\Gamma_{dt}^{num}$ with respect to $\epsilon$ allows to distinguish the deterministic and sweep-dominant regimes: in certain areas of the figures, the dynamic bifurcation threshold does not depend strongly on $\epsilon$, this is the DReg., while in other areas the dynamic bifurcation threshold depends on $\epsilon$, this is the SDReg.. The lower is the precision, the larger is the value of $\epsilon$ for which the DReg. appears. For each regime the theoretical results $\Gamma_{dt}^{th}$ and $\hat{\Gamma}_{dt}^{th}$ provides a good estimation of the dynamic oscillation threshold of the clarinet model~\eqref{dynsys_pp_exp}.

\section{Linear vs. exponential variation of the mouth pressure}
\label{sec:4}

\begin{figure}[h!]
\centering
\includegraphics[width=0.98\columnwidth]{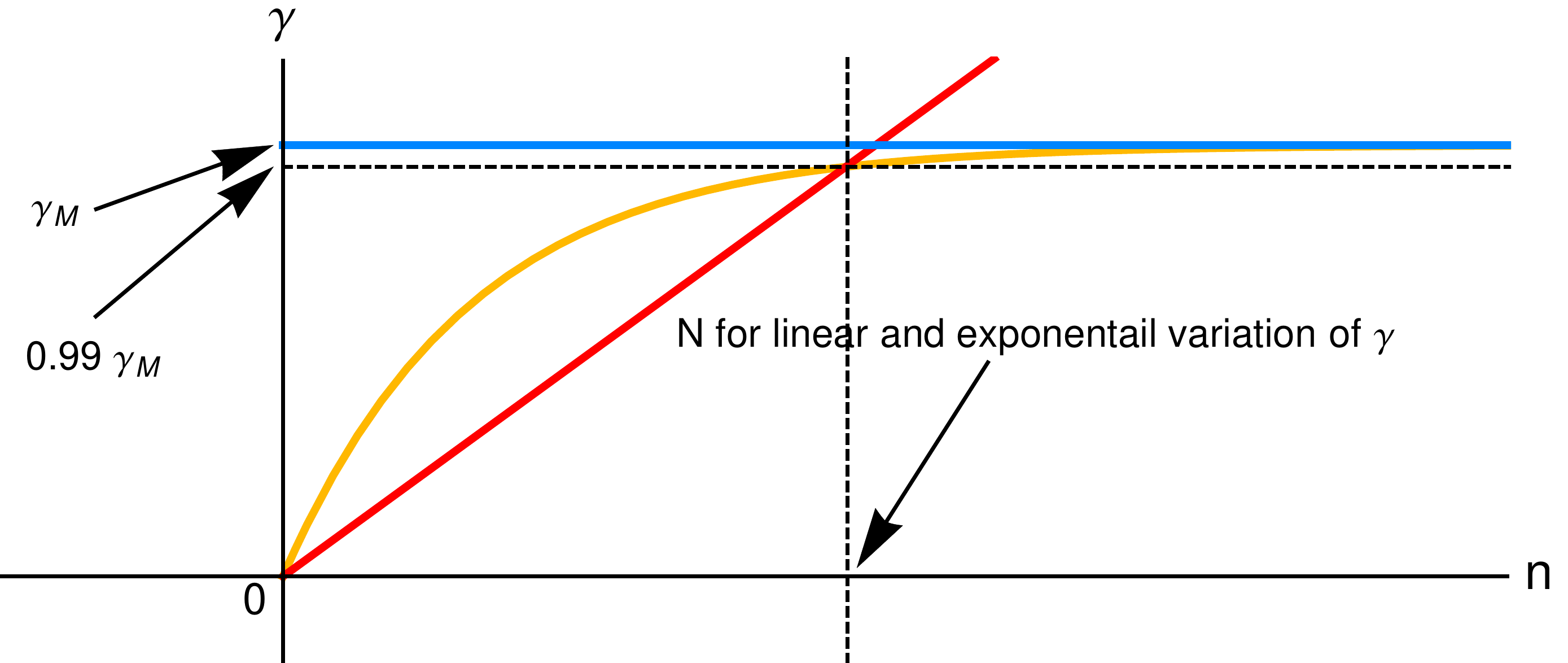}
\caption{Outline schematic showing the definition of $N$ for a linear and for an exponential variation of the mouth pressure $\gamma$.}
\label{fig:N}
\end{figure}

In this section, numerical dynamic threshold obtained for an exponential variation of the mouth pressure ($\Gamma_{dt}^{num}$) is compared to numerical dynamic threshold obtained for a linear variation of the mouth pressure ($\gamma_{dt}^{num}$). One can raise the question of the chosen parameter to carry out this comparison. As a preliminary investigation, we choose here the time (noted $N$) needed to reach 99 per cent of the target value $\gamma_M$ (see Fig~\ref{fig:N}).

For a linear variation of the mouth pressure with an increase rate $\epsilon$, we have: 
\begin{equation}
N=0.99\frac{\gamma_M}{\epsilon}.
\end{equation}

Inverting Eq.~\eqref{dynsys_pp_expfun} gives:

\begin{equation}
n=-\frac{1}{\epsilon}\ln\left[1-\frac{\gamma}{\gamma_M}\right].
\label{eq:nexp}
\end{equation}

Therefore, for an exponential variation of the mouth pressure, $N$ is defined by:

\begin{equation}
N=-\frac{1}{\epsilon}\ln\left[1-\frac{0.99\gamma_M}{\gamma_M}\right]\approx\frac{4.6}{\epsilon}.
\end{equation}

To illustrate the definition of the parameter $N$, Fig.~\ref{fig:N} shows a linear and an exponential functions plotted with the same value of $N$.

\begin{figure}[h!]
\centering
\includegraphics[width=1.03\columnwidth]{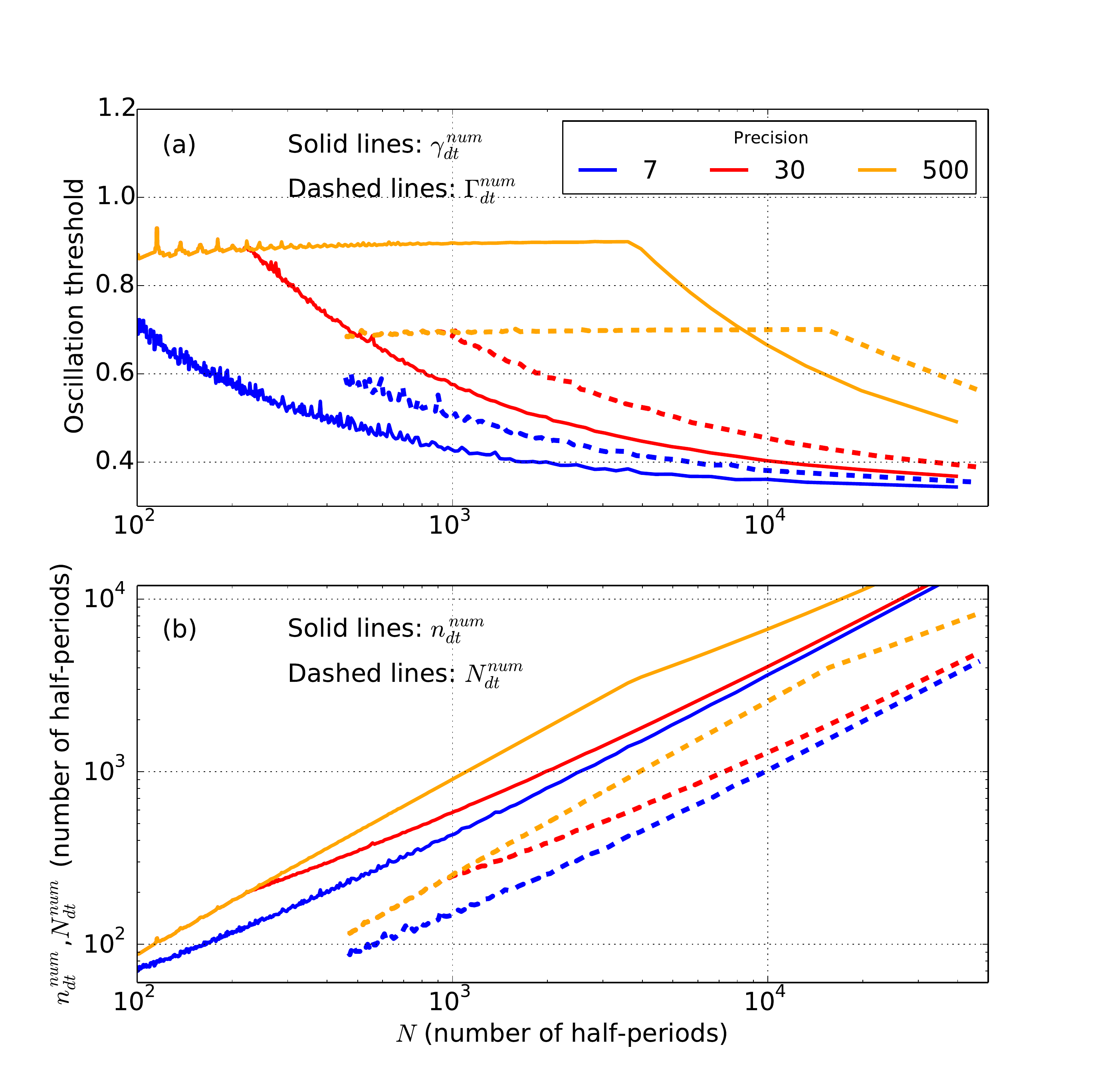}
\caption{Comparison between linear and exponential variation of the blowing pressure with respect to $N$. Different precisions are used: prec.~= 7, 30, and 500. (a) Comparison between the dynamic thresholds: $\Gamma_{dt}^{num}$ and $\gamma_{dt}^{num}$. (b) Comparison between the times to reach the dynamic thresholds: $N_{dt}^{num}$ and $n_{dt}^{num}$.}
\label{fig:2}
\end{figure}

The comparison is depicted in Fig.~\ref{fig:2}. Fig.~\ref{fig:2}(a) compares  the dynamic oscillation thresholds $\Gamma_{dt}^{num}$ and $\gamma_{dt}^{num}$: for the DReg., $\gamma_{dt}^{num}$ is  larger than $\Gamma_{dt}^{num}$ while, for SDReg., the opposite is noticed.

To complete the study, it is also interesting to compare the times $N_{dt}^{num}$ and $n_{dt}^{num}$ (computed through Eq.~\eqref{eq:nexp}) needed to reach $\Gamma_{dt}^{num}$ and $\gamma_{dt}^{num}$ (see Fig.~\ref{fig:2}(b)). For DReg. as well as for SDReg., exponential shape appears to provide faster onset of oscillations.


%
%

\section{Conclusion}


The method presented in this article provides an extension of the estimation of dynamic thresholds for exponentially increasing parameters. The present method can, in principle, be used for any other profile of time-varying parameter that can be described analytically.

Previous works have shown that, for sufficient precision, the dynamic threshold is independent of the rate of variation of the parameter (Deterministic regimes). A quick extrapolation of these results might have led to the conclusion that in a generic profile, even if there is a variation of the rate of change of $\gamma$, there would be no significant changes in the dynamic threshold. However the results in this article show that this is not the case, as the deterministic regime has a threshold (approximately 0.7) that is smaller than that of the linearly increasing profile (approximately 0.9).


In real cases, however, the system is always far from a deterministic regime, as the numerical or turbulence noise introduces a stochastic variation in the parameter then brings the system into the sweep-dominant regime. In these cases  (similarly to the deterministic regime), exponential shape appears to provide faster onset of oscillations. Obviously, additional works must be performed to state definitive conclusions about the influence of the mouth pressure shape on the onset of oscillations in a clarinet.


\begin{thebibliography}{10}

\bibitem{BergeotNLD2012}
B.~Bergeot, C.~Vergez, A.~Almeida, and B.~Gazengel.
\newblock Prediction of the dynamic oscillation threshold in a clarinet model
  with a linearly increasing blowing pressure.
\newblock {\em Nonlinear Dynam.}, 73(1-2):521--534, 2013.

\bibitem{BergeotNLD2012b}
B.~Bergeot, C.~Vergez, A.~Almeida, and B.~Gazengel.
\newblock Prediction of the dynamic oscillation threshold in a clarinet model
  with a linearly increasing blowing pressure: Influence of noise.
\newblock {\em Nonlinear Dynam.}, 74(3):591--605, 2013.

\bibitem{KergoActa2000}
J.~Kergomard, S.~Ollivier, and J.~Gilbert.
\newblock Calculation of the spectrum of self-sustained oscillators using a
  variable troncation method.
\newblock {\em Acta. Acust. united Ac.}, 86:665--703, 2000.

\bibitem{dalmont:3294}
J.~P. Dalmont, J.~Gilbert, J.~Kergomard, and S.~Ollivier.
\newblock An analytical prediction of the oscillation and extinction thresholds
  of a clarinet.
\newblock {\em J. Acoust. Soc. Am.}, 118(5):3294--3305, 2005.

\bibitem{Fruchard2007}
A.~Fruchard and R.~Sch{\"a}fke.
\newblock Sur le retard {\`a} la bifurcation.
\newblock In {\em International conference in honor of claude Lobry}, 2007.

\bibitem{Jasa2013BBergeot}
B.~Bergeot, A.~Almeida, B.~Gazengel, C.~Vergez, and D.~Ferrand.
\newblock Response of an artificially blown clarinet to different blowing
  pressure profiles.
\newblock {\em J. Acoust. Soc. Am.}, 135(1):479--490, 2014.

\bibitem{HirschAal1990}
A.~Hirschberg, R.~W. A.~Van de~Laar, J.~P. Maurires, A.~P.~J Wijnands, H.~J.
  Dane, S.~G. Kruijswijk, and A.~J.~M. Houtsma.
\newblock A quasi-stationary model of air flow in the reed channel of
  single-reed woodwind instruments.
\newblock {\em Acustica}, 70:146--154, 1990.

\bibitem{MOMIchap7}
A.~Hirschberg.
\newblock Aero-acoustics of wind instruments.
\newblock In {\em Mechanics of musical instruments by A. Hirschberg/ J.
  Kergomard/ G. Weinreich}, volume 335 of \textit{CISM Courses and lectures},
  chapter~7, pages 291--361. Springer-Verlag, 1995.

\bibitem{Maga1986}
C.~Maganza, R.~Causs{\'e}, and F.~Lalo{\"e}.
\newblock Bifurcations, period doublings and chaos in clarinet-like systems.
\newblock {\em EPL (Europhysics Letters)}, 1(6):295, 1986.

\bibitem{MechOfMusInst}
J.~Kergomard.
\newblock Elementary considerations on reed-instrument oscillations.
\newblock In {\em Mechanics of musical instruments by A. Hirschberg/ J.
  Kergomard/ G. Weinreich}, volume 335 of \textit{CISM Courses and lectures},
  chapter~6, pages 229--290. Springer-Verlag, 1995.

\bibitem{OllivActAc2005}
S.~Ollivier, J.~P. Dalmont, and J.~Kergomard.
\newblock Idealized models of reed woodwinds. part 2 : On the stability of
  two-step oscillations.
\newblock {\em Acta. Acust. united Ac.}, 91:166--179, 2005.

\bibitem{Cha08Belin}
A.~Chaigne and J.~Kergomard.
\newblock Instruments {\`a} anche.
\newblock In {\em Acoustique des instruments de musique}, chapter~9, pages
  400--468. Belin, 2008.

\bibitem{NonLin_Tail_2010}
P.A. Taillard, J.~Kergomard, and F.~Lalo{\"e}.
\newblock Iterated maps for clarinet-like systems.
\newblock {\em Nonlinear Dynam.}, 62:253--271, 2010.

\bibitem{Feigen1979}
M.~J. Feigenbaum.
\newblock The universal metric properties of nonlinear transformations.
\newblock {\em J. Stat. Phy.}, 21(6):669--706, 1979.

\bibitem{Baesens1991}
C.~Baesens.
\newblock Slow sweep through a period-doubling cascade: Delayed bifurcations
  and renormalisation.
\newblock {\em Physica D}, 53:319--375, 1991.

\bibitem{FruchaScaf2003}
A.~Fruchard and R.~Sch{\"a}fke.
\newblock Bifurcation delay and difference equations.
\newblock {\em Nonlinearity}, 16:2199--2220, 2003.

\end{thebibliography}

\end{document}